\documentclass[10pt,letterpaper]{article}
\usepackage[top=0.85in,left=2.75in,footskip=0.75in]{geometry}

\usepackage{amsmath,amssymb}

\usepackage{changepage}

\usepackage[utf8x]{inputenc}

\usepackage{textcomp,marvosym}

\usepackage{cite}

\usepackage{nameref,hyperref}

\usepackage{microtype}
\usepackage[table]{xcolor}

\usepackage{array}

\newcolumntype{+}{!{\vrule width 2pt}}

\newlength\savedwidth

\raggedright
\usepackage[aboveskip=1pt,labelfont=bf,labelsep=period,justification=raggedright,singlelinecheck=off]{caption}

\makeatletter
\renewcommand{\@biblabel}[1]{\quad#1.}
\makeatother

\usepackage{lastpage,fancyhdr,graphicx}
\usepackage{epstopdf}
\fancyheadoffset[L]{2.25in}
\fancyfootoffset[L]{2.25in}
\begin{document}
\vspace*{0.2in}

% Title must be 250 characters or less.
\begin{flushleft}
{\Large
%\textbf\newline{The structure of beyond the pairwise gene-gene interactions} % Please use "sentence case" for title and headings (capitalize only the first word in a title (or heading), the first word in a subtitle (or subheading), and any proper nouns).
\textbf\newline{The structure of gene-gene networks beyond pairwise interactions} % Please use "sentence case" for title and headings (capitalize only the first word in a title (or heading), the first word in a subtitle (or subheading), and any proper nouns).
}
\newline
% Insert author names, affiliations and corresponding author email (do not include titles, positions, or degrees).
\\
Nastaran Allahyari,
Amir Kargaran,
Ali Hosseiny,
G. R. Jafari*
%with the Lorem Ipsum Consortium\textsuperscript{\textpilcrow}
\\
\bigskip
Department of Physics, Shahid Beheshti University, Evin, Tehran 19839, Iran
\\
\bigskip

% Insert additional author notes using the symbols described below. Insert symbol callouts after author names as necessary.
% 
% Remove or comment out the author notes below if they aren't used.
%
% Primary Equal Contribution Note
%\Yinyang These authors contributed equally to this work.

% Additional Equal Contribution Note
% Also use this double-dagger symbol for special authorship notes, such as senior authorship.
%\ddag These authors also contributed equally to this work.

% Current address notes
%\textcurrency Current Address: Dept/Program/Center, Institution Name, City, State, Country % change symbol to "\textcurrency a" if more than one current address note
% \textcurrency b Insert second current address 
% \textcurrency c Insert third current address

% Deceased author note
%\dag Deceased

% Group/Consortium Author Note
%\textpilcrow Membership list can be found in the Acknowledgments section.

% Use the asterisk to denote corresponding authorship and provide email address in note below.
* g\_jafari@sbu.ac.ir

\end{flushleft}
% Please keep the abstract below 300 words
\section*{Abstract}

Despite its high and direct impact on nearly all biological processes, the underlying structure of gene-gene interaction networks is investigated so far according to pair connections. To address this, we explore the gene interaction networks of the yeast Saccharomyces cerevisiae beyond pairwise interaction using the structural balance theory (SBT). Specifically, we ask whether essential and nonessential gene interaction networks are structurally balanced. We study triadic interactions in the weighted signed undirected gene networks and observe that balanced and unbalanced triads are over and underrepresented in both networks, thus beautifully in line with the strong notion of balance. Moreover, we note that the energy distribution of triads is significantly different in both essential and nonessential networks compared with the shuffled networks. Yet, this difference is greater in the essential network regarding the frequency as well as the energy of triads. Additionally, results demonstrate that triads in the essential gene network are more interconnected through sharing common links, while in the nonessential network they tend to be isolated. Last but not least, we investigate the contribution of all-length signed walks and its impact on the degree of balance. Our findings reveal that interestingly when considering longer cycles the nonessential gene network is more balanced compared to the essential network.

%\linenumbers

\section*{Introduction}

Today, various studies investigate genomic information based on pairwise connections in gene interaction networks \cite{Barabasi}. However, the interesting collective behaviors that emerge from these interactions can not be described by simply considering pairs of genes. In other words, while studying pair connections has well broadened our view on the functionality of genes, the higher-order organizations are yet to be explored. To be specific, studies demonstrate that genes are categorized into two main groups \cite{Zhou}. Functionally, essential genes play a more vital role in the biological process, and locally they form a denser network compared to nonessential genes. Yet the crucial question raised here is if there exists a structure beyond these pairwise interactions in these two networks? If so, what is the difference in the underlying structure between essential and nonessential networks? Suppose in a signed interaction network genes $A$, $B$, and $C$ are connected, is it logical to consider the interaction AB detached from its context, that is, triad ABC? What is the impact of interactions $AC$ and $BC$ on the interaction between genes $A$ and $B$? It is known that triadic interactions play a significant role in the construction of real-world networks \cite{Leskovec, Belaza}, and structural balance theory (SBT) has well discussed these interactions. In this work, we apply SBT to the gene interaction networks to answer the following questions: Is there a structure beyond pairwise interaction in the gene interaction networks? Which types of triads, balanced or unbalanced, are over (under) represented in these networks compared to the shuffled networks regarding both the frequency and the energy distributions? Is there a difference between essential and nonessential networks in the pattern of connection between triads? In addition, when considering all lengths of cycles, which network is more balanced? And do all genes have an equal impact on the final networks’ degree of balance? These questions are the basis of this study.

SBT was introduced in social psychology by Heider to investigate the structure of tension in networks whose mutual relationships are explained in terms of friendship and hostility \cite{Heider_new}. Later this theory has been generalized for graphs by Cartwright and Harary through considering the triads as low-dimensional motifs \cite{Cartwright_new}. One of the standard applications provided by balance theory is to measure the degree of balance/ stability in networks \cite{Easley, Kirkley, Aref, Khanafiah, Harary2, Norman}. On the other hand, quantifying the degree of unbalance/ frustration in a signed network was proposed as well \cite{Facchetti}. Similarly, in biological networks distance to the exact balance is computed \cite{Iacono, Dasgupta, Maayan, Sontag}. Moreover, several researchers have studied the dynamics based on which an unbalanced network achieves balance through reducing unbalanced triads\cite{antal1, antal2, Marvel1, Marvel2, Abell, Traag, Gawronski, Hedayatifar}. Some studies provide further theoretical expansion of balance theory employing methods from Boltzmann-Gibbs statistical physics to unravel the dynamics behind the structural balance \cite{Belaza, Rabbani, Castellano}. An appealing application of balance theory recently applied predicts which correlation matrix coefficients are likely to change their signs in the high-dimensional regime \cite{borjiorno}. Consequently, there have been two main trends in the literature of SBT: 1) Studying the analytical aspects theoretically \cite {antal2, Kulakowski, Krawczyk, harrary, Rijt, Kargaran, Masoumi, Oloomi}, 2) Applying it to a wide variety of real-signed social, economic, ecologic, and political networks empirically to clarify their structures\cite{Hart, Hummon, Doreian, Lerner, Saiz, Moradimanesh, Zheng, Ilany}. Amongst these applications, it should be mentioned that understanding the structure entirely, not partially, calls for considering not only short-range interactions but also longer-range cycles\cite{Davis, Acharya, Srinivasan, Estrada}. Accordingly, we analyze the structural balance of gene interaction networks. We study the genetic interaction profile similarity matrices of the yeast Saccharomyces cerevisiae\cite{Costanzo, Costanzo2}, which has been categorized into two main classes, namely, essential and nonessential. Amongst all 5500 genes, approximately 1000 genes are essential because of their vital functional role in biological processes. According to the threshold taken by Costanzo and et al. in \cite{Costanzo}, essential genes have higher degrees and are considered hubs in the global network. Thus these genes play a considerable role in the local structure of the network. On top of that, essential genes have higher prediction power compared to nonessential genes \cite{Blomenand, Winzeler}.

Here, we investigate the weighted, signed, and undirected networks of genetic interaction for essential and nonessential genes of the yeast Saccharomyces cerevisiae. Primarily, we are interested in probing the existence of structure beyond the pairwise gene interactions in these networks. To this aim as in our previous study \cite{Allahyari}, we compare the spectrum of eigenvalues between genetic interaction matrices and their shuffled versions. The rest of the paper is organized as follows. First, we explore the frequency of triads in the gene networks according to the notion of over and under representation of different types of triad compared to the shuffled networks. Afterward, we assign energy levels to unique configurations of triads and demonstrate triads’ energy distributions. Then, the energy-energy mixing patterns between triads are analyzed to systematically investigate how triads with different energies are connected in the networks. Additionally, we examine the balance of the gene interaction networks by considering all lengths of cycles. Last but not least, we propose a list of significant genes which have a high impact on the global degree of networks’ balance.

\section*{Materials and methods}

\begin{figure}[h]{
		\includegraphics[height=10.5cm,width=1.\linewidth]{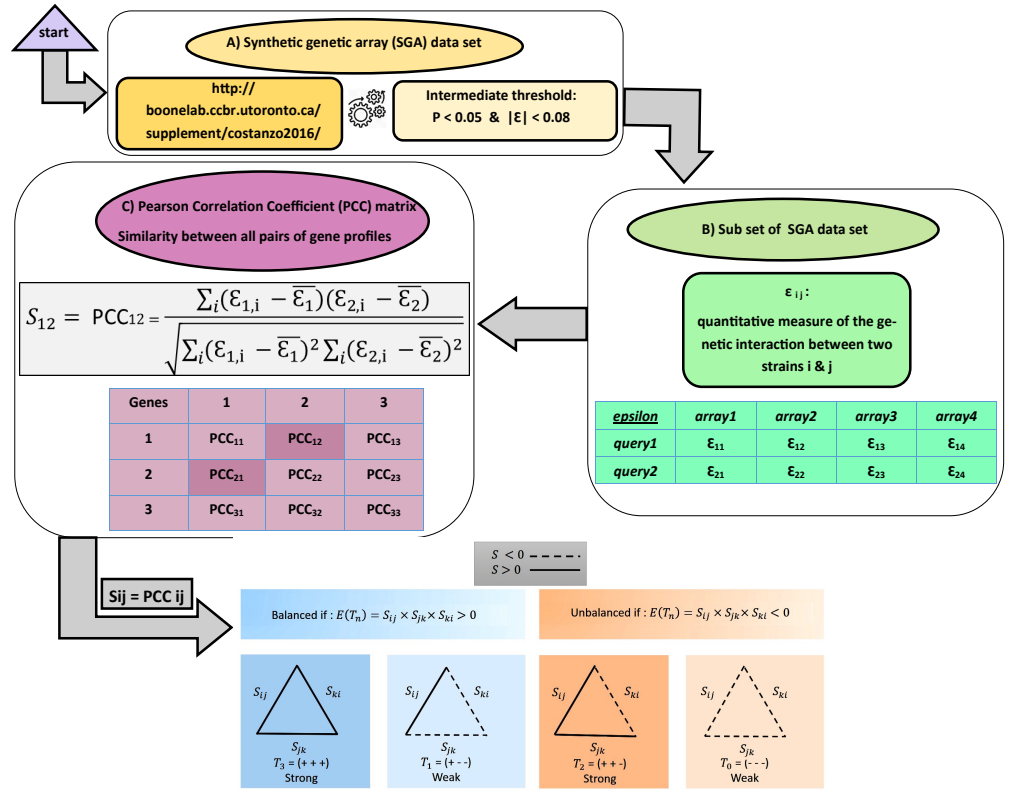}
	}
	\caption{\bf Graphical abstract for the procedure of obtaining the genetic interaction similarity matrices.}
	\label{fig1}
\end{figure}

{\bf Data.} {\it Saccharomyces cerevisiae} is a beneficial yeast to analyze eukaryotes. One of the outstanding characteristics of it is that almost all bioprocesses in eukaryotes can exist in Saccharomyces cerevisiae \cite{Parapouli}. Here, we analyze the gene interaction similarity networks of about 5500 genes. Around 1000 genes are identified as essential, and the rest of them as nonessential genes \cite{Baryshnikova, Li}. Costanzo and his colleagues have provided the data\cite{Costanzo}.  They have published three gene interaction similarity matrices ,for essential genes, nonessential genes and the combination of them in the global form. The data file used is available at \url{http://boonelab.ccbr.utoronto.ca/supplement/costanzo2016/}. We have worked with data file S3 titled "Genetic interaction profile similarity matrices". The steps taken by them to produce this data are as below:
\\
A) Based on the growth rate of the colony consisting of two specific mutated genes, the genetic interaction score (epsilon) between them has been obtained.
\\
B) A genetic interaction profile for each gene is constructed by considering the genetic interaction score between that gene and a set of other genes in the colony.
\\
C) The similarity between all two profiles is obtained by calculating the Pearson correlation coefficient (PCC).
\\
The positive value in the PCC matrix indicates how much those two genes are functionally similar to each other and vise versa. Moreover, zero elements show that those two genes are not related to each other functionally. We represent the procedure accomplished to obtain the PCC matrices in Fig \ref{fig1}.

{\bf Network analysis.} Before anything else, to understand the networks working with, some features were calculated. The network's topological and statistical measurements analyzed here are: Mean degree, the ratio between mean of squared degrees and squared of mean degree, modularity, assortativity coefficient, average path length, and clustering coefficient. The coefficient $\langle k^2 \rangle$ holds information about the values around mean degree. However, $\langle k \rangle^2$ includes information about the tail of degree distribution. Hence, low $\frac{\langle k^2 \rangle}{\langle k \rangle^2}$ indicates that the tail carries a higher share in the couplings. About modularity, it measures the strength of a network in division into modules. As another feature, assortativity (positive coefficient) means that a high-degree component usually prefers to be connected with the high-degree one and vice versa \cite{newman}. Disassortativity (negative coefficient) implies that a giant cluster tends to link with a small one. Also, mean length declares that, on average, how nodes can create a relationship with each other \cite{Albert}. Finally, a high clustering coefficient states the extent to which the agents in the system tend to remain in their clusters \cite{Watts}.

After analyzing network features, it is substantial to examine if the network construction is random or not. So, the existence of structure beyond the pairwise interaction in the gene interaction network is analyzed. When there is no structure beyond pairwise interactions, that network can be known as a random one. In a random network, the distribution of the spectrum of eigenvalues has a semi-circular form with a body-centered around zero \cite{Jalan}. In a nonrandom network, there are some eigenvalues out of the bulk\cite{Pradhan}. Also, one large eigenvalue exists that mostly has a value far from the bulk of the eigenvalues\cite{Namaki, Saeedian}. This eigenvalue plays a significant role and addresses the global trend of the system.

{\bf Structural balance theory.} To go beyond the assumption that pair interactions are independent and looking for triads as the shortest motif, structural balance theory (SBT) is applied\cite{Kulakowski}. To consider the local triangles, we focus on groups with three interacting genes in the network. There are four kinds of triads, including two balanced and two unbalanced ones. The idea of "The friend of my friend is also my friend" [+ + +]  refers to a strongly balanced triad.  The idea of "The enemy of my enemy is my friend" [− − +] points to a weakly balanced triad. The two other types of signed triadic configurations, [+ + −] is a strongly unbalanced triad, and [− − −] is a weakly unbalanced triad. These triads give rise to frustration in the network\cite{Davis}. In other words, the triangle is recognized as a balanced one if the sign of the product of its links is positive. Otherwise, the triangle is unbalanced or frustrated. Significant computational methods are used to speed up accounting for the number of triads in the signed and large network\cite{Terzi}. It works based on connectivity (G) and adjacency (A) matrices. In the connectivity matrix, $G(i,j) = 1$ if the nodes $i$ and $j$ are connected, otherwise $G(i,j) = 0$. In the adjacency matrix, $A(i,j) = 1$ represents all positive elements in graph and the $A(i,j) = -1$  denotes all negative interactions in the graph. As below, the two equations count the number of balanced $b$ and frustrated $u$  triads, respectively:

\begin{equation}
b=\frac{1}{12}[trace(G^3)+trace(A^3)],
\label{eq1}
\end{equation}

\begin{equation}
u=\frac{1}{12}[trace(G^3)-trace(A^3)].
\label{eq2}
\end{equation}

As Leskovek has proposed \cite{Leskovec}, we have built a null model to compare the empirical frequencies of triads.  It is important for generating a null model to keep the exact fraction of positive (negative) signs. Each selected link is randomly connecting the two existing nodes.  So the created null model represents no organization in the structure. Then, we calculate the fraction of each kind of triad in the shuffled network as $p_{0}(T_{i})$.  The triad $i$ is overrepresented if the related fraction in the original network as $p(T_{i})$ be more than that of the shuffled one. Otherwise, it will be underrepresented. Next, the value of surprise, as bellow is calculated in which $T_{i}$ is the number of triad $i$ and $E[T_{i}]$ is the expected number of triad $i$ calculated as $E[T_{i}]= {\Delta}p_{0}(T_{i})$ and ${\Delta}$ is the total number of triads and $p_{0}(T_{i})$ as mentioned before is the fraction of triad $i$ in the shuffled network. To eliminate the effect of size in both networks, after calculating the $s(T_{i})$ function, it is divided into $\sqrt{\Delta}$.

\begin{equation}
s(T_{i}) = \frac{T_{i}-E[T_{i}]}{\sqrt{{\Delta}p_{0}(T_{i})(1-p_{0}(T_{i}))}}.
\label{eq33}
\end{equation}

It has been stated that a balanced network is a network consisting of all positive triads\cite{Harary2}. While the possibility of possessing a real-world network containing all positive signed triads (positive product of their sides) is close to zero. So a common approach is to measure the degree of balance of a signed network. To this aim, the concept of balance enables us to determine an energy landscape for such networks. Energy describes how much a network is structurally balanced. The network energy is obtained by the negative summation of the products of the triads' links ($S_{ij}S_{jk}S_{ki}$) divided by the total number of triads (${\Delta}$)\cite{Marvel2,somaye}. If the network energy (${E}$) is $-1$, then we have a fully balanced network. But if it equals $+1$, then we will have an unbalanced network. Consequently, in real-world networks, the energy of triads is between $-1$ and $+1$. According to SBT's suggestion, a network evolves towards the minimum level of tension between triadic \cite{somaye}.

\begin{equation}
E = - \frac{1}{\Delta}\sum_{i<j<k}^N S_{ij}S_{jk}S_{ki}.
\label{eq3}
\end{equation}

The energy landscape introduced above considers the triads individually and does not designate how they are connected. The energy-energy mixing pattern between triads shows which of them with energy $E_{1}$ has a common link with the other one with energy $E_{2}$. So we can find out that concerning the energy value, what triangles are contiguous to connect. This pattern shows if the specific types of triangles are packed together and form a kind of module. Also, this pattern figures out if the triads represented a heterogeneous pattern of connections. Moreover, triangles with higher energies prefer to connect to ones with lower or the same value.

{\bf The walk-based measure of balance and detecting lack of balance.} SBT gives informative information to understand the structural balance of signed networks but is biased. Through these small groups, our analysis recognizes the frustration on the shortest possible cycle, but it overlooks to considering the unbalance correlated with longer-range cycles\cite{Kargaran}. Being a balanced or unbalanced cycle is related to the multiplication of the signs of its edges. If the sign of the product is negative, or the number of negative links in the cycle is odd, it is an unbalanced cycle. If all of them in a network has a positive sign, we can consider the signed network as a balanced one \cite{Davis, Acharya, Srinivasan}. The probability of having a network with real data containing all cycles with a positive sign is close to zero. As Estrada proposed in \cite{Estrada}, we calculate the walk-balance index $(K)$ for walks with all lengths by assigning more weights to the shorter ones, which is logical \cite{Estrada}. This method relates a hypothetical equilibrium between the real-world signed network and its underlying unsigned version:

\begin{equation}
K=\frac{trace\Big(exp\big(A(\Sigma)\big)\Big)}{trace\Big(exp\big(A(| \Sigma |)\big)\Big)}.
\label{eq4}
\end{equation}

Where $A(\Sigma)$ and $A(| \Sigma |)$ are signed and unsigned adjacency matrices, respectively. Elements in $A(\Sigma)$ are $+1$ when the interaction matrix values are more than zero. Also, if the interaction matrix values are less than zero the elements in $A(\Sigma)$ are $-1$. In the unsigned adjacency matrix $A(| \Sigma |)$, if the elements in the interaction matrix are nonzero, the elements of $A(| \Sigma |)$ are $1$. Another index that can measure the extent of the lack of balance in the network $(U)$ is as follows \cite{Estrada}:

\begin{equation}
U=\frac{1-K}{1+K}.
\label{eq5}
\end{equation}

When a network is highly unbalanced, $K \approx 0$, it implies $U \approx 1$. Diversely, a balanced network has K = 1 and U = 0. At last, the participation of each node in the balance of the network can be calculated by the degree of balance of a given node $i$ as $K_{i}$ \cite{Estrada}:

\begin{equation}
K_{i}=\frac{exp\big( A(\Sigma)\big)_{ii}}{exp\big( A(|\Sigma|)\big)_{ii}}.
\label{eq6}
\end{equation}

\section*{Results}

\begin{figure}[h]{
		\includegraphics[height=6.4cm,width=.9\linewidth]{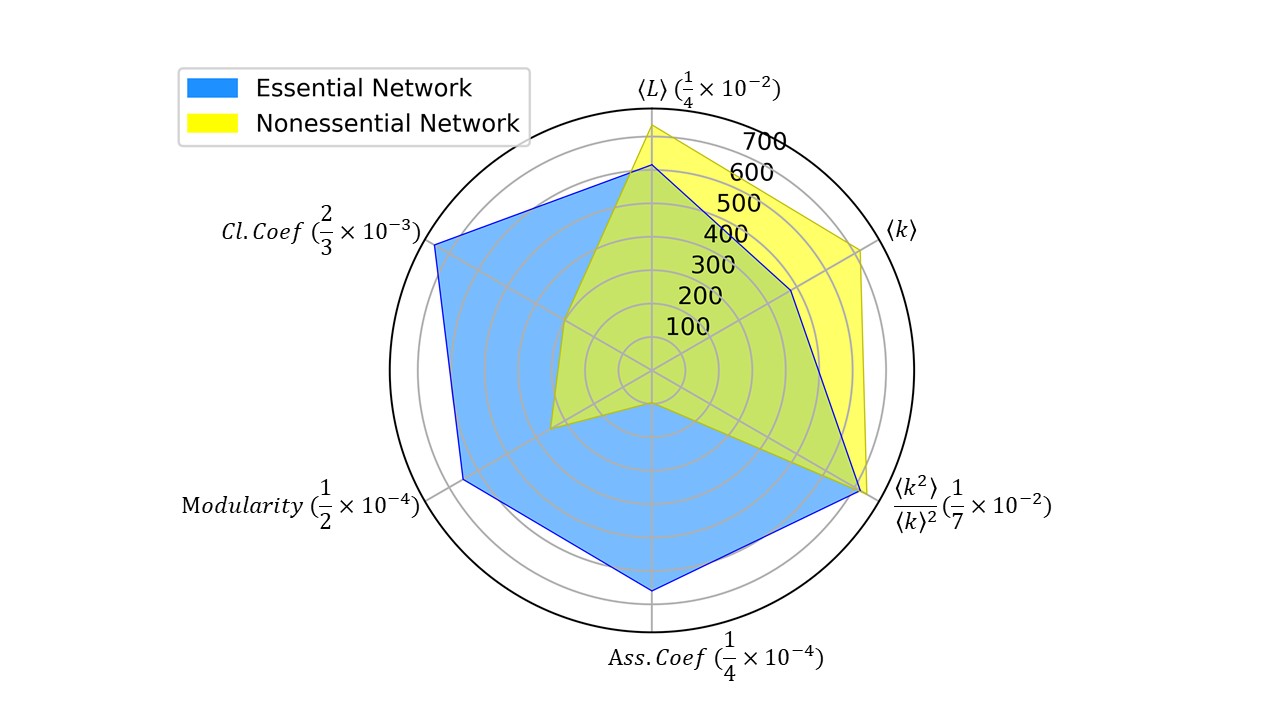}
	}
	\caption{{\bf The radar plot shows both essential and nonessential gene networks' features including: Mean degree, the ratio between mean of squared degrees and squared of mean degree, modularity, assortativity coefficient, average path length, and cluster coefficient.} The radar plot for the essential gene network is plotted in blue and the nonessential gene network in yellow.}
	\label{fig2}
\end{figure}

First, important features in both essential and nonessential gene networks are compared in Fig \ref{fig2}. Despite the segregation among the measurements, there exist some similarities. As shown in Fig \ref{fig2}, the mean degree $\langle k \rangle$ in nonessential gene network is higher compared to the essential network. Besides, in both networks, the ratio between mean squared degrees and squared of mean degree is close to one. This implies that neither nodes with high degree nor medium degree are significantly dominant over the other one. In addition, both networks have nearly similar modularity, as a measure of a network's tendency to cluster into multiple sets of strongly interacting parts, with a little higher degree for the essential gene network. Moreover, as it has been illustrated in Table \ref{table1}, the assortativity coefficient in both networks is negative but so close to zero. That is, both networks show weak disassortative behavior. However, the magnitude of assortativity is one order higher in the essential network. In the radar plot (Fig \ref{fig2}), we demonstrated the absolute values of assortativity coefficients. The other significant feature of the networks is the average path length which represents the number of steps along the shortest path for each pair of nodes. The small value of this characteristic in both networks shows that these networks are densely connected, however for nonessential networks it is a bit longer. At last, the tendency in forming clusters as defined by the clustering coefficient is higher in the essential network.

\begin{table}[h]
	\caption{Network's features: Mean degree, the ratio between mean of squared degrees and squared of mean degree, modularity, assortativity coefficient, average path length, and cluster coefficient for both essential and nonessential gene networks.
	}
	\label{table1}
	\centering
	\renewcommand{\arraystretch}{1.2}
	\begin{tabular}{|c||c|c|c|cc}
		\rowcolor{gray}
		\hline
		Gene  Networks   &   Essential     & Nonessential   & Proportion of Features $\frac{Nonessential}{Essential}$  \\
		\hline
		\hline
		$\langle k \rangle$  & $478.890$&  $718.957$ & $1.501$ \\
		\hline
		$\frac{\langle k^2 \rangle}{\langle k \rangle^2}$ & $1.028$&  $1.060$ & $1.0313$ \\
		\hline
		$Modularity$ & $0.033$&  $0.018$ & $0.536$ \\
		\hline
		$Assortativity$ & $-0.017$ & $ -0.002$ & $0.148$ \\
		\hline
		$\langle L \rangle$ & $1.539$& $1.838$ & $1.194$ \\
		\hline
		$Clustering.Coef$ & $0.501$& $0.201$  & $0.402$ \\
		\hline

	\end{tabular}
\end{table}

\begin{figure}[h]{
		A) \includegraphics[height=5.1cm,width=.45\linewidth]{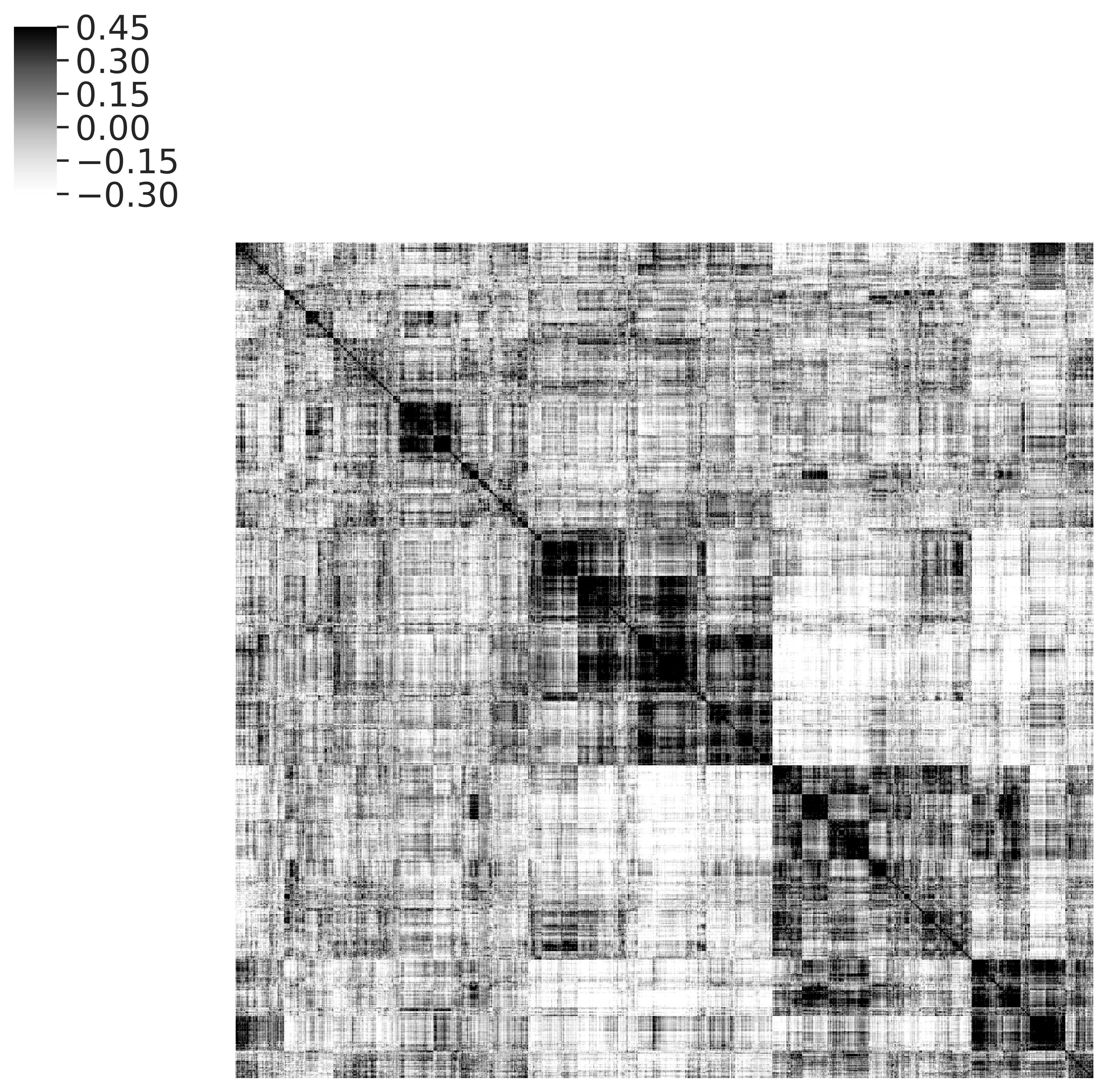}
		B) \includegraphics[height=4.cm,width=.35\linewidth]{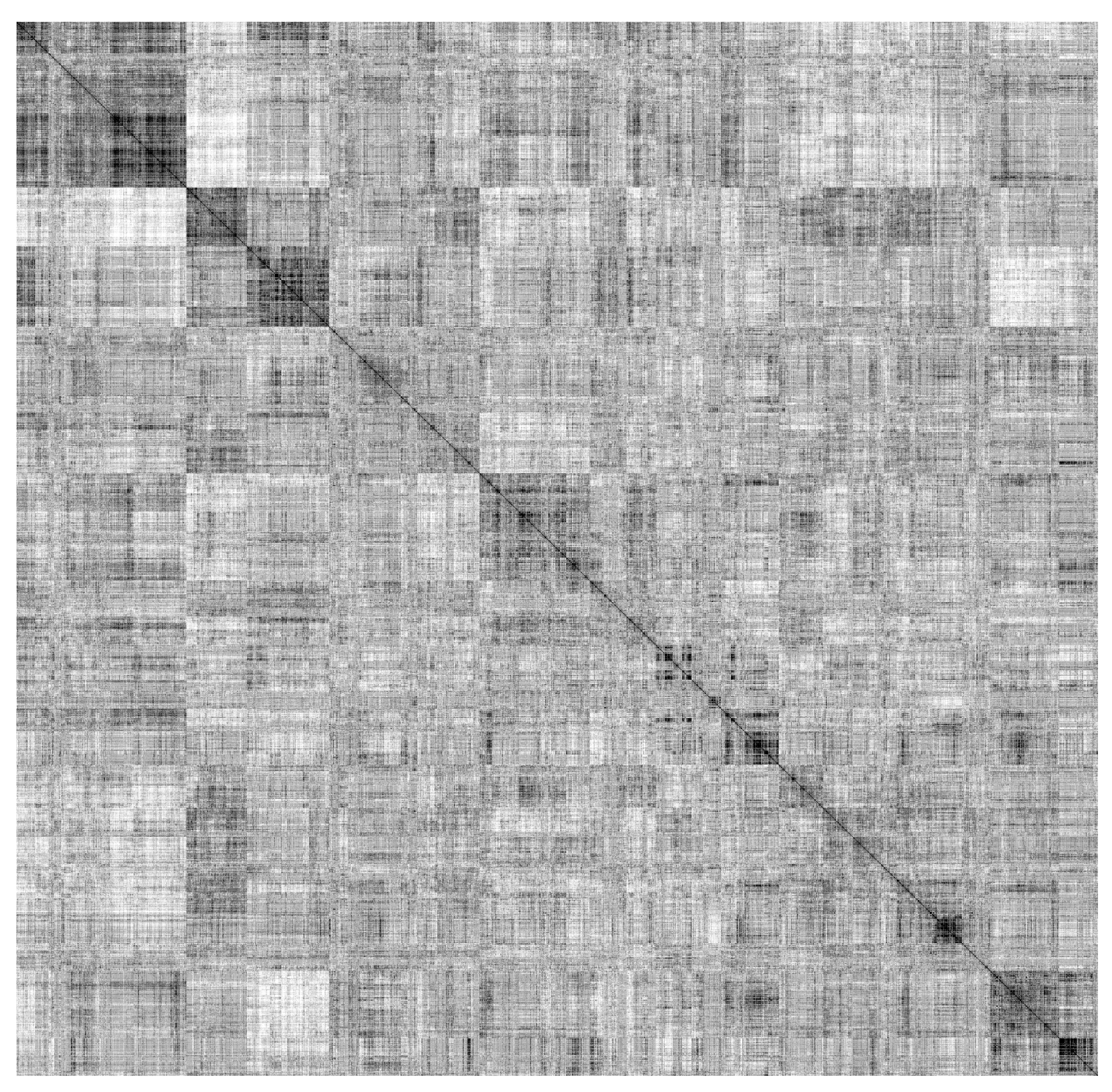}
	}
	\caption{{\bf The cluster map of two essential and nonessential gene networks.} A) Cluster map of essential gene network, B) Cluster map of nonessential gene network.}
	\label{fig3}
\end{figure}

Then, we have investigated the existence of clusters in the construction of the essential and nonessential gene networks. Within groups, the genes cooperate to annotate a common bioprocess efficiently. Clusters in the essential and nonessential gene networks are illustrated in cluster maps (Fig \ref{fig3}). It can be seen that the essential network has higher modularity which is in line with the previous result which stated that the essential network is more densely connected than the nonessential network. In other words, although the clusters exist in both networks, the structure in the essential gene network (Fig \ref{fig3}A) is highly stronger than the nonessential network (Fig \ref{fig3}B). This is also confirmed in our previous work, where we observed a significant difference between the distributions of eigenvalues in original matrices and the shuffled networks \cite{Allahyari}. To be specific, some of the eigenvalues in the original networks are not limited to the narrow bulk of the shuffled matrices' eigenvalues. Thus, it can be confidently concluded that the structure of the gene interaction networks is far from random.

The structural balance in gene interaction networks to study the structure beyond pairwise interactions is analyzed. In Table \ref{table2}, the size, and the percentage of positive and negative links, and the total number of triads in both networks are prepared. In the following, the two equations Eq \eqref{eq1} and Eq \eqref{eq2} are used to count balanced $b$ and unbalanced $u$ triads. To compare the dominance of balanced or unbalanced triads in our networks, we have used a method proposed by Leskovec et al.\cite{Leskovec}. If the fraction of a triad in the original network is higher than the shuffled one, it will overrepresent, and vise versa. The fraction of the triad $T_{i}$ in the original network is considered as $p(T_{i})$ and in the shuffled network $p_{0}(T_{i})$. Moreover, they have proposed the concept of surprise as Eq \eqref{eq1}, $s(T_{i})$, to understand how significant these over (under) representations are. Due to the size of the networks, $s(T_{i})$ has a significant order of tens. Balanced triads are overrepresented in both essential and nonessential gene interaction networks. On the contrary, unbalanced triads are underrepresented compared to the shuffled. These results are presented in Table \ref{table3}.

\begin{table}[h]
	\caption{{\bf Dataset statistics.} Number of nodes, edges, triads in both essential and nonessential gene networks with threshold $S_{ij}<|0.05|$.}
	\label{table2}
	\centering
	\renewcommand{\arraystretch}{1.2}
	\begin{tabular}{|c||c|c|c}
		\rowcolor{gray}
		\hline
		Gene networks        &  Essential  & Nonessential     \\
		\hline
		\hline
		$ Nodes $  &  $1,040$   & $4,430$  \\
		\hline
		 $ Edges $&  $249,023$   & $1,592,490$  \\
		\hline
		$ + $ $ Edges $ &  $50.1\%$   & $63.5\%$  \\
		\hline
		$ - $ $ Edges $ &  $49.9\%$   & $36.4\%$  \\
		\hline
		$\frac{Edges}{\binom N2}$&  $0.461$   & $0.162$  \\ 	\hline
		$ Triads $& $20,310,741$   & $81,470,554$  \\
		\hline
		$\frac{Triads}{\binom N3}$& $0.109$   & $0.006$  \\ \hline
	\end{tabular}
\end{table}

\begin{table}[!h]
	\caption{{\bf Number and probability of balanced and unbalanced triads in the original networks compared to the null model.} $|T_{i}|$, the total number of $Ti$ ; $p(T_{i})$, the fraction of $T_{i}$; $p_{0}(T_{i})$, the fraction of $T_{i}$ in the null model; $s(T_{i})$, the amount of surprise, i.e., is the number of standard deviations by which the actual number of $T_{i}$ differs from its expected number under the null model.}
	\label{table3}
	\centering
	\begin{tabular}{|c||c|c|c|c|c|c}
		\rowcolor{gray}
		\hline
		Essential  gene network &   $|T_{i}|$     &  $p(T_{i})$  & $p_{0}(T)$  &  $s(T_{i})$ & $\frac{s(T_{i})}{\sqrt{\Delta}}$\\
		\hline
		$Strongly$ $balanced$ $(T_{3})$ & $3,670,948$  &  $0.180$ & $0.124$ & $764.00$ & $0.170$ \\
		\hline
		$Weakly$ $balanced$ $(T_{1})$ & $10,362,180$&  $0.510$  & $0.375$ & $1,255.11$ & $0.278$\\
		\hline
		$Strongly$ $unalanced$ $(T_{2})$ & $4,421,666$  &  $0.217$ & $0.374$ & $-1,461.09$ &$-0.324$ \\
		\hline
		$Weakly$ $unalanced$  $(T_{0})$ & $1,855,947$&  $0.091$  & $0.125$ & $-462.01$ & $-0.103$\\
		\hline
		\hline
		\hline
		\rowcolor{gray}
		Nonessential gene network  &     $|T_{i}|$     &  $p(T_{i})$  & $p_{0}(T)$  &  $s(T_{i})$  & $\frac{s(T_{i})}{\sqrt{\Delta}}$\\
		\hline
		$Strongly$ $balanced$ $(T_{3})$ & $30,868,604$  &  $0.378$ & $0.256$ & $2,531.12$ & $0.280$\\
		\hline
		$Weakly$ $balanced$  $(T_{1})$ & $32,704,022$&  $0.401$  & $0.253$ & $3,071.60$ & $0.340$\\
		\hline
		$Strongly$ $unalanced$ $(T_{2})$ & $16,028,365$  &  $0.196$ & $0.441$ & $-4,452.75$ &$-0.493$ \\
		\hline
		$Weakly$ $unalanced$ $(T_{0})$ & $1,869,563$&  $0.022$  & $0.048$ & $-1,071.69$ & $-0.119$\\
		\hline
	\end{tabular}
\end{table}

\begin{figure}[!h]{
	    \flushleft
		A)\includegraphics[height=3.6cm,width=.4\linewidth]{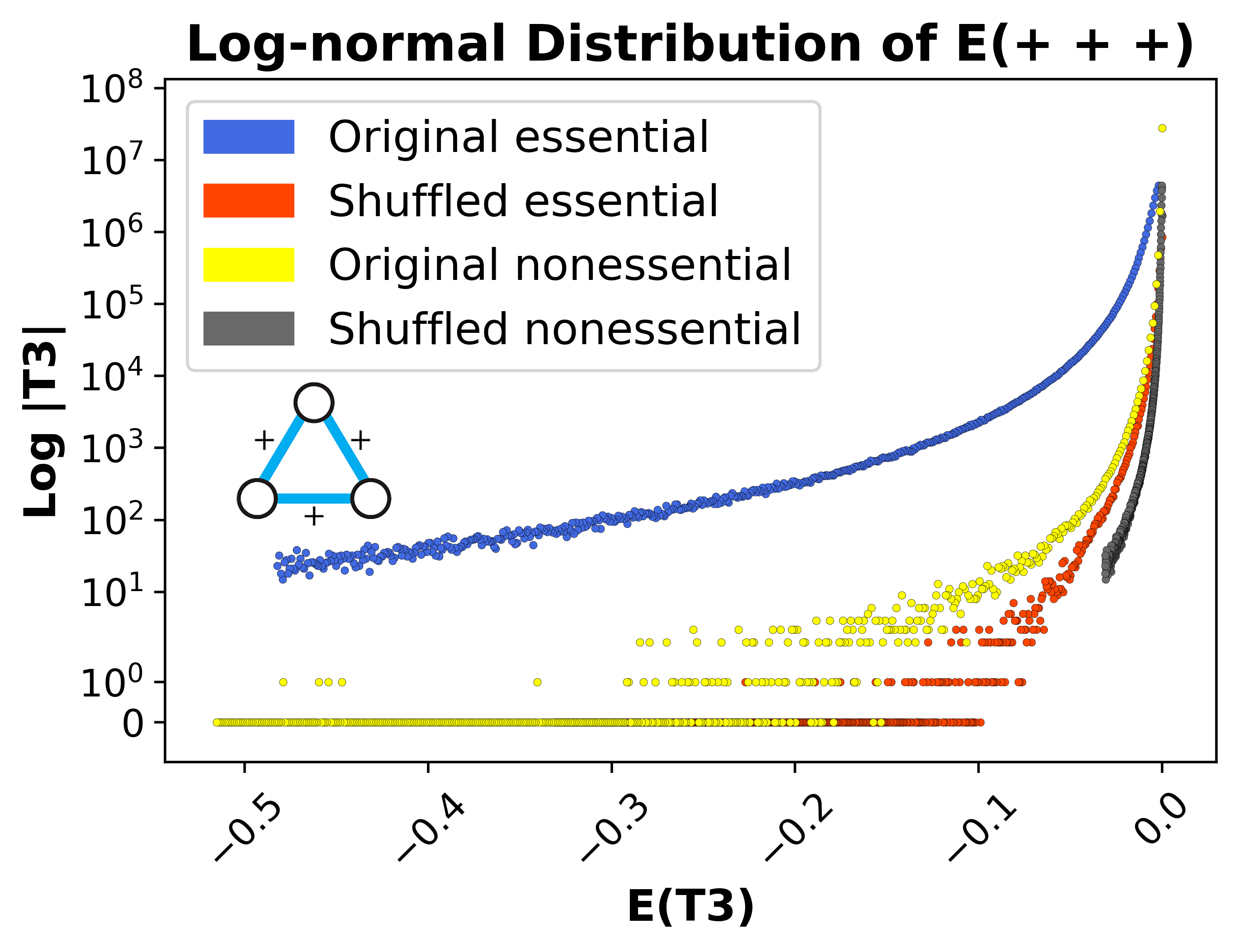}
		B)\includegraphics[height=3.6cm,width=.4\linewidth]{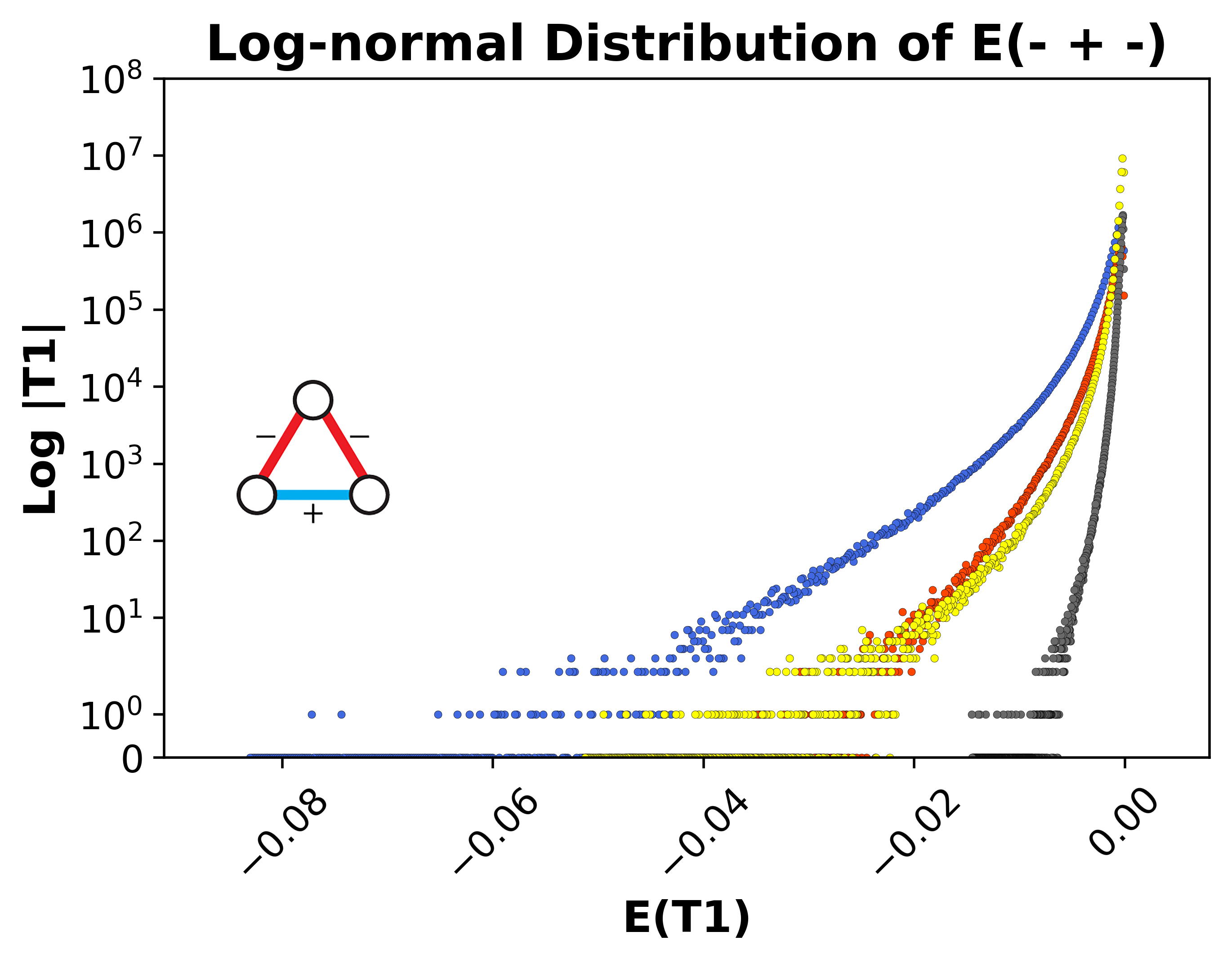}\\
		C)\includegraphics[height=3.6cm,width=.4\linewidth]{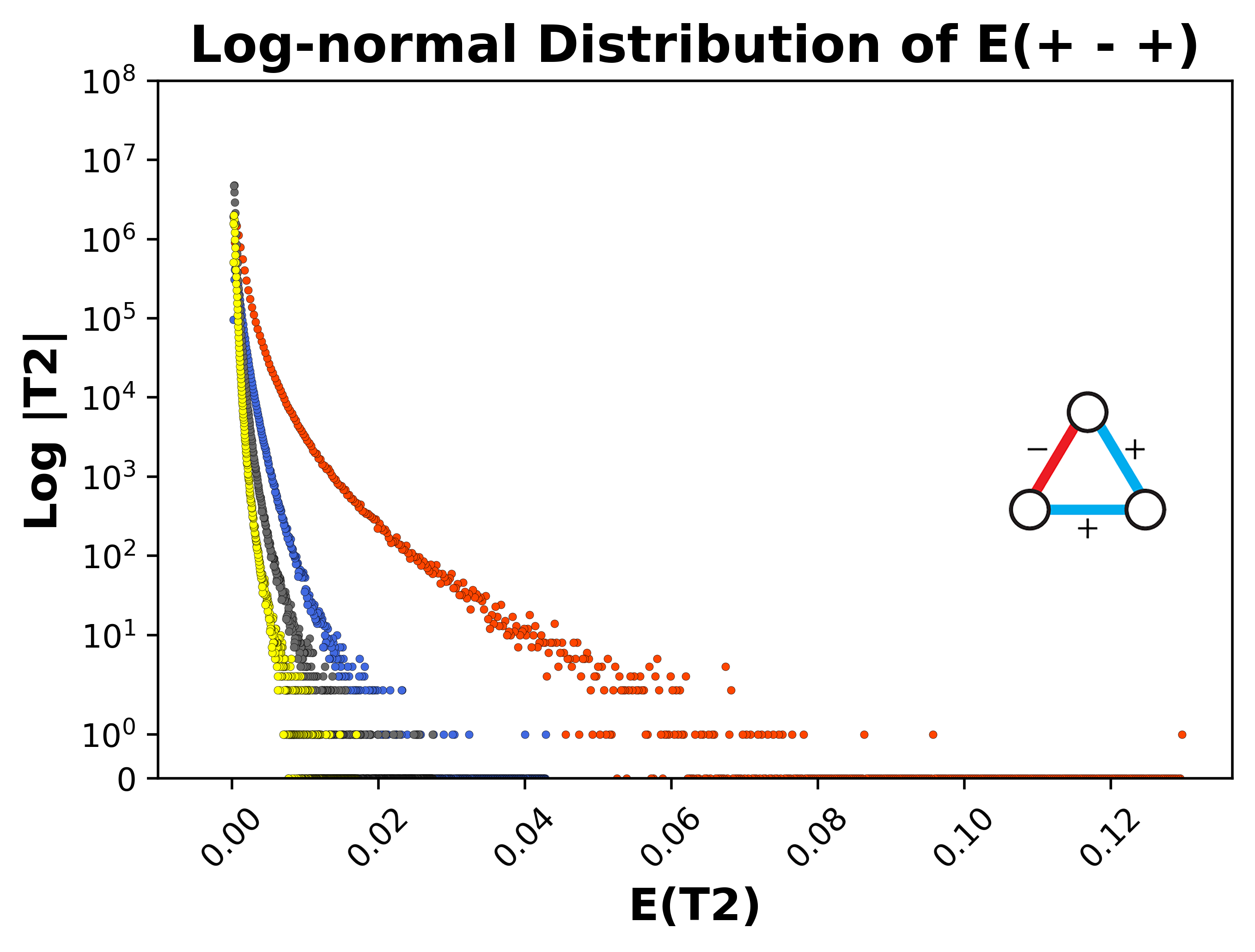}
		D)\includegraphics[height=3.6cm,width=.4\linewidth]{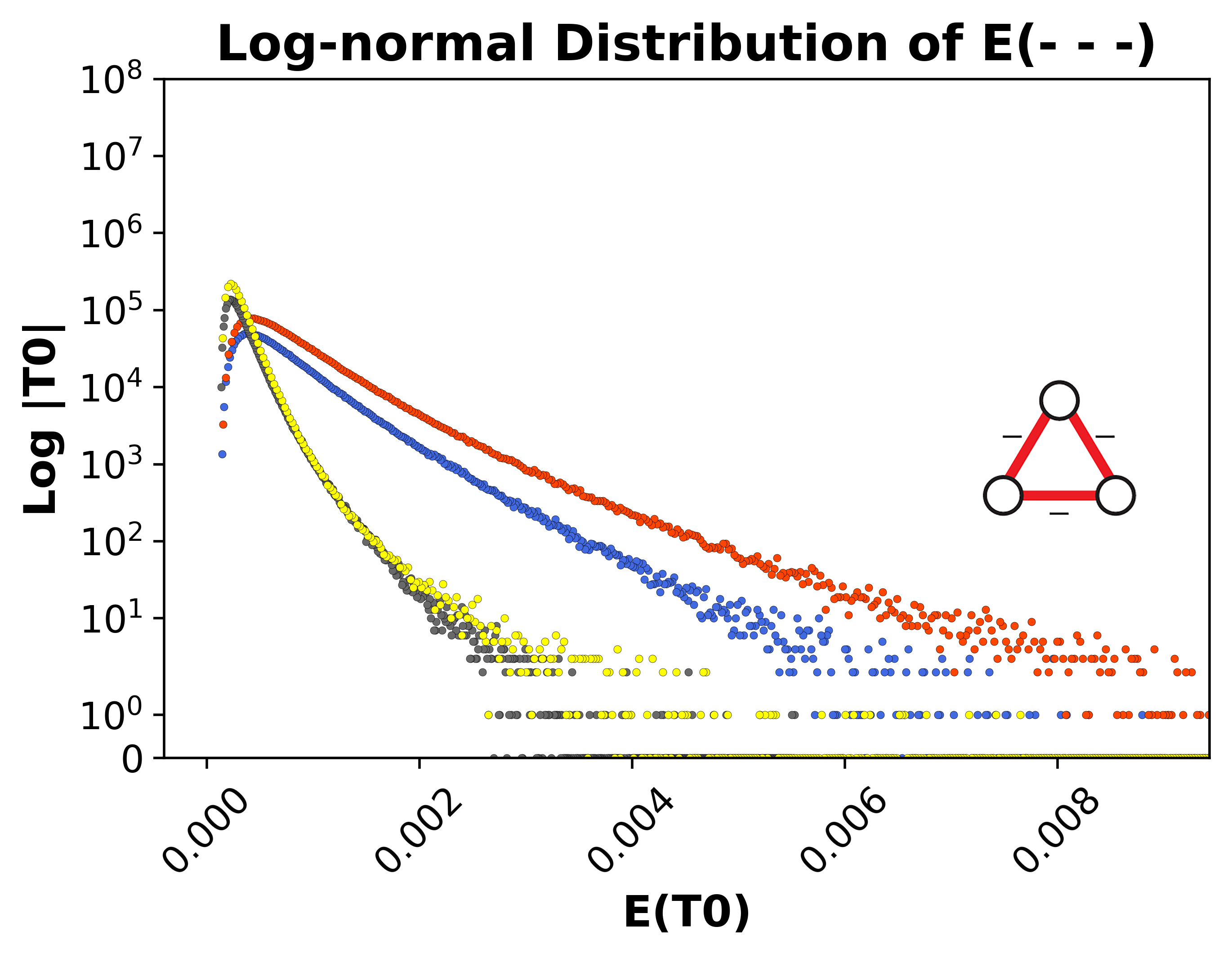}}\\
    	\flushleft
    	E)\includegraphics[height=3.6cm,width=.4\linewidth]{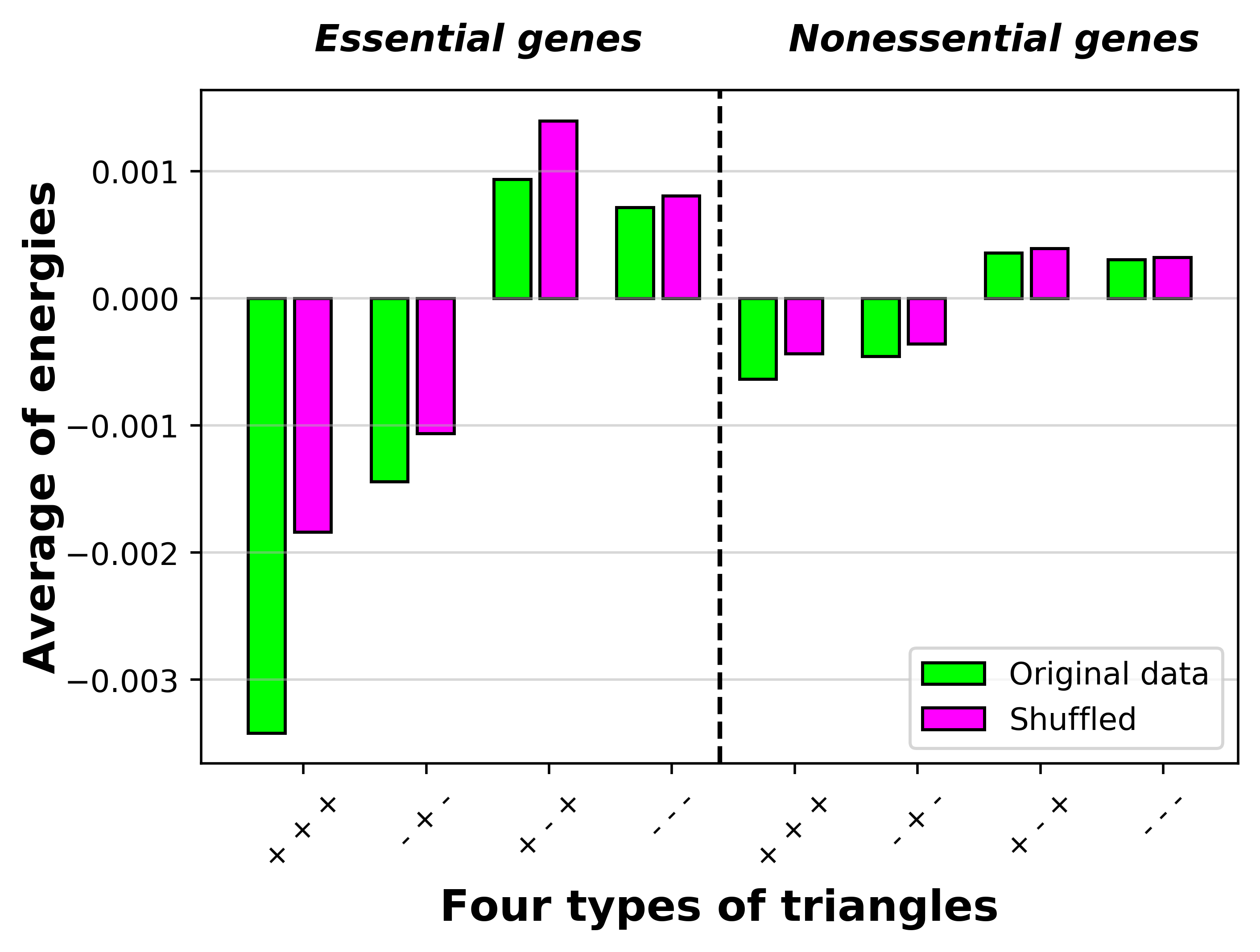}
    	F)\includegraphics[height=3.6cm,width=.4\linewidth]{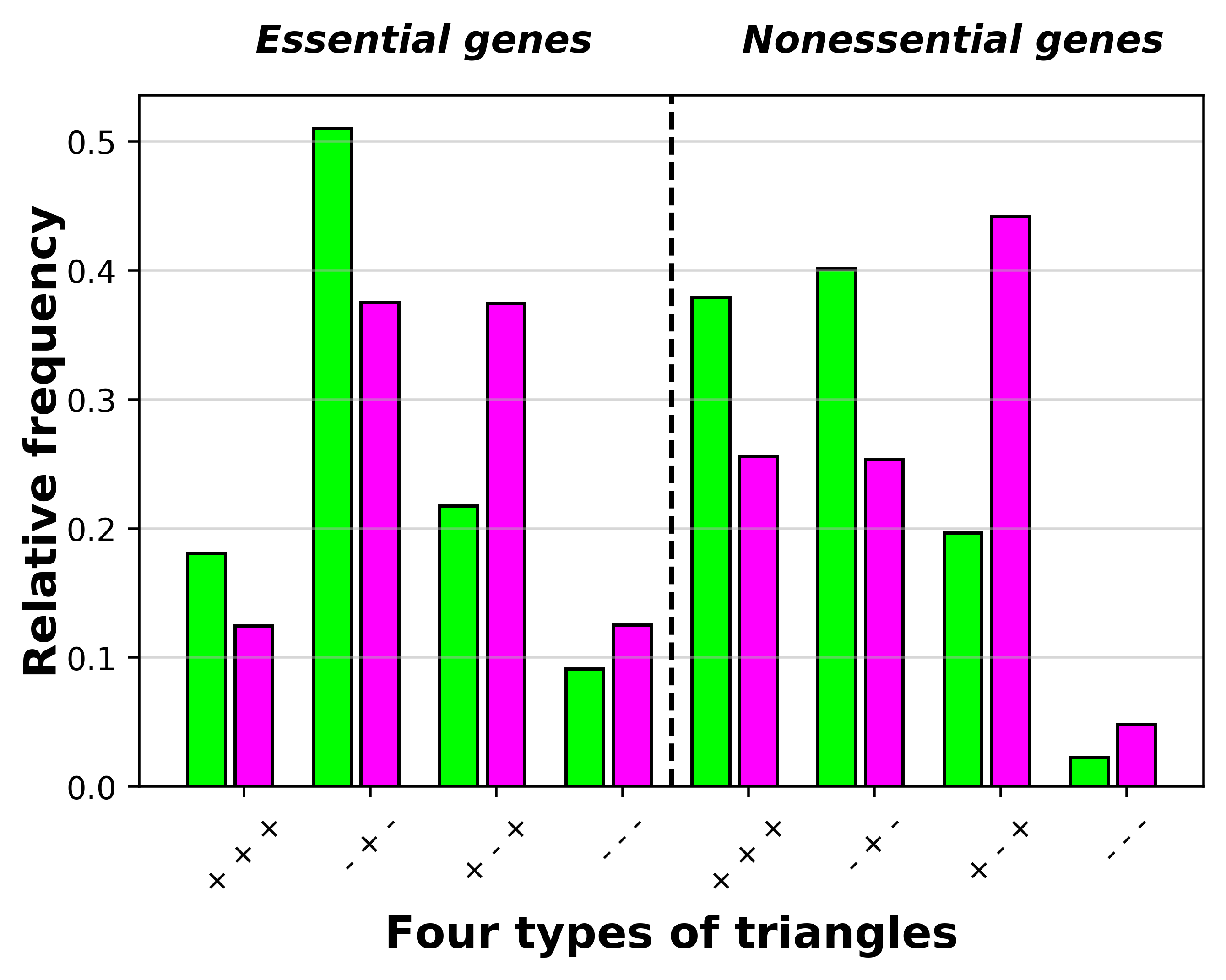}
	\caption{{\bf Energy distributions for all four types of triads in log scale.} A) Energy distribution for strongly balanced triads, B) Energy distribution for weakly balanced triads, C) Energy distribution for strongly unbalanced triads, D) Energy distribution for weakly unbalanced triads. (The energy distribution of triads for original essential gene network and its shuffled network are plotted in blue and red, respectively. The energy distribution of triads for original nonessential gene network and its shuffled network are plotted in yellow and gray, respectively.) {\bf The average energy for all four kinds of triangles.} E) From left to right, essential gene network and nonessential gene network. {\bf The relative frequency for all four kinds of triangles.} F) From left to right, essential gene network and nonessential gene network (Green bars for original networks and purple ones for shuffled networks.)}
	\label{fig4}
\end{figure}

After analyzing the frequency of triads, we have examined the energy distribution of different types of triads. So we have calculated the energy of triads by Eq \eqref{eq3}. Then the energy distributions of strongly balanced triads in Fig \ref{fig4}A, weakly balanced triads in Fig \ref{fig4}B, strongly unbalanced triads in Fig \ref{fig4}C, and weakly unbalanced triads in Fig  \ref{fig4}D for both original networks, in comparison with their shuffled, are presented. Results indicate: 1) All kinds of triads, in both essential and nonessential networks, have many triads with small energies. 2) In the essential gene network as Fig \ref{fig4}E, the average energy of all types of triads is larger than the nonessential triads. 3) In the essential gene network, like the nonessential network, the bar levels of the average energy of balanced triads are higher than shuffled ones. However, on the contrary, the bar levels of the average energy of unbalanced triads are lower than shuffled ones. 4) As Fig \ref{fig4}F, in the essential gene network, the bar level of the relative frequency of the balanced triad with one positive side is individually equal to the relative frequency of the other three triads.

\begin{figure}[h]{
		\includegraphics[height=6.6cm,width=.9\linewidth]{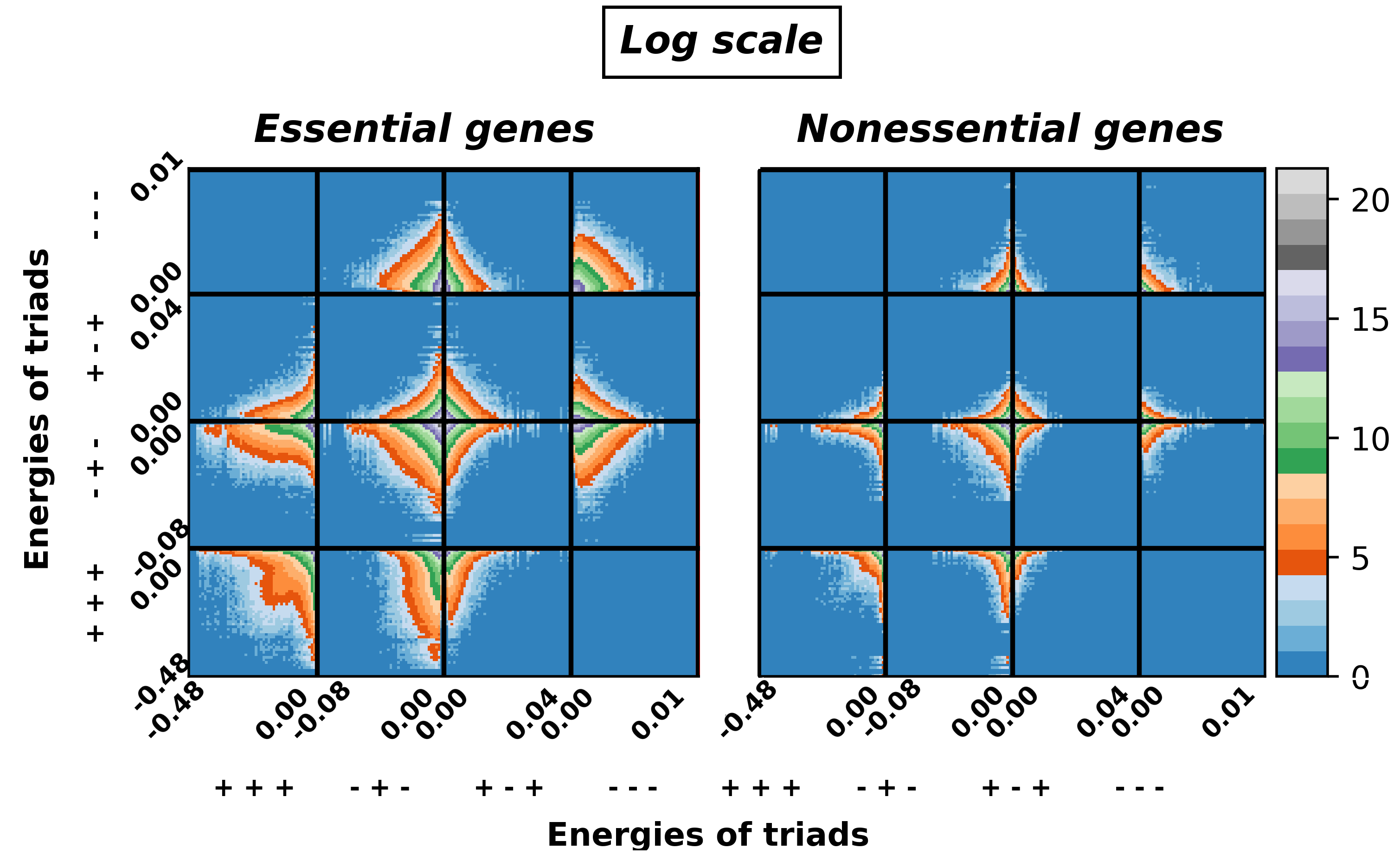}
	}
	\caption{{\bf The pattern of connection between different types of triads with  different energies through one shared link for both original networks in Log scale.} From left to right, essential gene network and nonessential gene network}
	\label{fig5}
\end{figure}

Through another consideration, we look for triads with one shared link in the networks. We display the energy-energy mixing pattern between the triangles. Fig \ref{fig5} shows how many triangles with different energies are connected. To have a more accurate consideration, the logarithmic scale of that analysis has been plotted. By using the logarithmic scale, there is a magnification between the elements with small amounts. The same behavior from both networks is observed. This plot reflects more sparsity for one shared link in the nonessential gene network rather than the essential gene network. However, the essential gene network shows more preference to participate in modules than nonessential genes. This result is notable because the number of triads in the nonessential gene network is much more than that of the essential gene network. Moreover, the triads with higher absolute valued energies have a shared edge with higher magnitude energies.

Now, by considering walks with all possible lengths, we extend our analysis. The quantity of balance or unbalance through these walks is measured. Indeed, we used the two indices introduced in \cite{Estrada} by Estradato not to limit ourselves only to triads as the shortest cycle. The walk-balance index by Eq \eqref{eq4} defines the quantification of how close to balance an unbalanced network is. Another index represents the amount of the shortage of balance in a given signed interaction network by Eq \eqref{eq5}. In Table \ref{table4}, the amounts of the walk-balance index in both essential and nonessential gene networks have been presented. The result indicates that by considering all walks, the nonessential gene network is more close to balance.  Besides, the extent of the lack of balance in the essential gene network is much more than the nonessential gene network. Also, we shuffled the interaction matrices and calculated these indexes again. There is a leading difference for each index between the result of the original and the shuffled matrices. Furthermore, there is an index that characterizes the degree of balance for a given node by Eq \eqref{eq6}. In supplementary, a Table is prepared to represent the classification of highlighted essential and nonessential genes based on their significant role in the balance.

\begin{table}[h]
	\caption{Walk-balance index for all cycles (K), Percentage of the lack of balance (U) in the original and shuffled of essential and nonessential gene networks with threshold $S_{ij}<|0.2|$.}
	\label{table4}
	\centering
	\begin{tabular}{|c||c|c|cc}
		\rowcolor{gray}
		\hline
		Gene  Networks   &   Essential     & Nonessential     \\
		\hline
		\hline
		$ K_{original\ network} $  & $0.195$&  $0.988$   \\
		\hline
		$K_{shuffled} $ & $0.000$&  $0.131$  \\
		\hline
		\hline
		$ U_{original\ network} $ & $67.238\%$ & $0.575\%$  \\
		\hline
		$ U_{shuffled} $ & $99.999\%$& $76.749\%$   \\
		\hline
	\end{tabular}
\end{table}

\newpage

\section*{Discussion}

We analyzed gene interactions in the weighted, undirected, and signed networks of yeast Saccharomyces cerevisiae. The pre-processed data set used includes two matrices, namely, essential and nonessential gene interaction networks. Here, we explored these two gene networks beyond pairwise interactions in the context of structural balance theory (SBT). The following results have been concluded accordingly: We have discovered that in both essential and nonessential gene networks balanced triads are overrepresented while unbalanced triads are underrepresented. Interestingly, this finding is in agreement with Heider’s balance theory. To be specific, our results empirically support the strong notion of structural balance theory (Table 2). This is while in some social networks, the weak formulation of structural balance has been reported as well \cite{Leskovec}.

Additionally, we have observed $T_{1}$ and $T_{0}$ triads in both gene networks with more average energy and higher relative frequency in the essential network. This can be interpreted from the perspective of SBT in which the presence of $T_{1}$ and $T_{0}$ triads in the organization of a network is related to having a higher degree of modularity. In other words, to have $T_{1}$ or $T_{0}$ in the stable state of a network indicates that densely connected modules are also connected to each other through negative links. This result corresponds to the presence of specialized clusters in the gene interaction network which has also been reflected in the energy-energy mixing pattern between the triads with one common link Fig \ref{fig5}. It is worth mentioning that this pattern is more significant in the essential network as genes in this network are more densely interconnected.

Moreover, we have noted that although energies of the essential and nonessential networks are not significantly different from each other, the underlying triads' distributions that led to these final energies are not similar. As mentioned earlier, the average energy and the relative frequency of unbalanced triads are higher in the essential gene network compared to the nonessential network. Thus, they are more likely to experience different possible states. Therefore, it can be concluded that unbalanced triads are providing the essential gene networks with the necessary structure that is needed to contain a dynamism which is crucial for vital biological mechanisms. This is while for nonessential genes with less unbalanced triads, the likelihood of being trapped in a local minima is higher.

Finally, to extend our analysis we have calculated two indices by considering the walks with all possible lengths. Namely, the quantification of how close to balance an unbalanced network is, and the extent to which a given signed network lacks balance by considering longer-range cycles. Results surprisingly suggest that when all length walks are taken into account, the nonessential gene network is more balanced and stable than the essential network. In other words, essential genes respect shorter-range connections while in nonessential genes long-range interactions have a higher impact. As mentioned earlier, the combination of both essential and nonessential interactions constructs the global gene network as a whole. For this network, we have proposed a list of genes that have an influential role in determining the final networks’ degree of balance. Thus, our finding highlights the genes that are structurally of note, regarding which further biological analysis seems to be very much valuable.

\section*{Author contributions}

{\bf Conceptualization:} Nastaran Allahyari, Gholam Reza Jafari, Ali Hosseyni.

{\bf Formal analysis:} Nastaran Allahyari, Gholam Reza Jafari, Ali Hosseyni, Amir Kargaran.

{\bf Methodology:} Nastaran Allahyari, Gholam Reza Jafari, Ali Hosseyni.

{\bf Resources:} Nastaran Allahyari, Amir Kargaran, Gholam Reza Jafari.

{\bf Software:} Nastaran Allahyari, Amir Kargaran.

{\bf Supervision:} Gholam Reza Jafari, Ali Hosseyni.

{\bf Visualization:} Nastaran Allahyari.

{\bf Writing:} Nastaran Allahyari.

\section*{Acknowledgments}
N.A. would like to express her appreciation to Z. Moradimanesh and S. Salekzamankhani for constructive comments that improved the manuscript.

%\nolinenumbers


\begin{thebibliography}{10}
\bibitem{Barabasi} Barabasi AL, Oltavi ZN. Network biology: understanding the cell's functional organization. Nature Reviews Genetics. 2004; 5(2):101-113. \url{https://doi.org/10.1038/nrg1272} PMID:\href{https://pubmed.ncbi.nlm.nih.gov/14735121/}{14735121}
\bibitem{Zhou} Zhou X, Menche J, Barabási AL, and Sharma A. Human symptoms disease network. Nature. Communications. 2014; 5, 1–10. \url{https://doi.org/10.1038/ncomms5212} PMID:\href{https://pubmed.ncbi.nlm.nih.gov/24967666/}{24967666}
\bibitem{Leskovec}  Leskovec J, Huttenlocher D, Kleinberg J. Signed networks in social media. Proc. SIGCHI Conf. Hum. Factors Comput. Syst. 2010; 22, 1361–1370. \url{https://doi.org/10.1145/1753326.1753532}
\bibitem{Belaza}  Belaza AM, Hoefman K, Ryckebusch J, Bramson A, Heuvel MVD, Schoors K. Statistical physics of balance theory. PLOS ONE. 2017; 12(8), e0183696. \url{https://doi.org/10.1371/journal.pone.0183696} PMID:\href{https://pubmed.ncbi.nlm.nih.gov/28846726/}{PMC5573279}
\bibitem{Heider_new} Heider F. Attitudes and Cognitive Organization.  Journal of Psychology. 1946; 21:107–112. \url{https://doi.org/10.1080/00223980.1946.9917275} PMID:\href{https://pubmed.ncbi.nlm.nih.gov/21010780/}{21010780}
\bibitem{Cartwright_new} Cartwright D, Harary F. Structural balance: a generalization of Heider’s theory. Psychological Review. 1956; 63(5):277–93. \url{https://doi.org/10.1037/h0046049} PMID: \href{https://pubmed.ncbi.nlm.nih.gov/13359597/}{13359597}
\bibitem{Easley} Easley D, Kleinberg J. Networks, Crowds, and Markets, Reasoning About a Highly Connected World. Cambridge University Press. 2010. \url{https://doi.org/10.1017/S0266466609990685}
\bibitem{Harary2} Harary F. On the measurement of structural balance. Behavioral Science. 1959; 4(4), 316–323. \url{https://doi.org/10.1002/bs.3830040405}
\bibitem{Khanafiah} Khanafiah D, Situngkir H. Social Balance Theory. arXiv:nlin/0405041 [Preprint]. 2014 [cited 2014 May 16]. Available from: \url{https://arxiv.org/abs/nlin/0405041}
\bibitem{Norman} Norman RZ, Roberts FS. A derivation of a measure of relative balance for social structures and a characterization of extensive ratio systems. Journal of Mathematical Psychology. 1972; 9(1), 66–91. \url{https://doi.org/10.1016/0022-2496(72)90006-5} 
\bibitem{Aref} Aref S, Wilson MC. Balance and frustration in signed networks. Journal of Complex Networks. 2018; 7(2), 163–189. \url{https://doi.org/10.1093/comnet/cny015}
\bibitem{Kirkley} Kirkley A, Cantwell GT, Newman MEJ. Balance in signed networks. Physical Review E. 2019; 99,012320. \url{https://doi.org/10.1103/PhysRevE.99.012320}
\bibitem{Facchetti} Facchetti G, Iacono G, Altafini C. Computing global structural balance in large-scale signed social networks. Proceedings of the National Academy of Sciences. 2011; 108(52):20953–20958. \url{https://doi.org/10.1073/pnas.1109521108}
\bibitem{Iacono} Iacono G, Ramezani F, Soranzo N, Altafini C. Determining the distance to monotonicity of a biological network: a graph-theoretical approach. IET Systems Biology. 2010; 4(3), 223–235. \url{https://doi.org/10.1049/iet-syb.2009.0040} PMID:\href{https://pubmed.ncbi.nlm.nih.gov/20500002/}{20500002}
\bibitem{Dasgupta} Dasgupta B, Enciso GA, Sontag E, Zhang Y. Algorithmic and complexity results for decompositions of biological networks into monotone subsystems. Biosystems. 2007; 90(1) 161– 178. \url{https://doi.org/10.1016/j.biosystems.2006.08.001} PMID:\href{https://pubmed.ncbi.nlm.nih.gov/17188805/}{17188805}
\bibitem{Maayan} Ma’ayan A, Lipshtat A, Iyengar R, Sontag ED. Proximity of intracellular regulatory networks to monotone systems. IET Syst. Biol. 2008; 2, 103 – 112. \url{https://doi.org/10.1049/iet-syb:20070036} PMID:\href{https://pubmed.ncbi.nlm.nih.gov/18537452/}{18537452}
\bibitem{Sontag} Sontag ED. Monotone and near-monotone biochemical networks. Systems and Synthetic Biology. 2007; 1, 59–87. \url{https://doi.org/10.1007/s11693-007-9005-9} PMID:\href{https://pubmed.ncbi.nlm.nih.gov/19003437/}{19003437}
\bibitem{antal1} Antal T, Krapivsky PL, Redner S. Dynamics of social balance on networks. Physical Review E. 2005; 72:036121. \url{https://doi.org/10.1103/PhysRevE.72.036121}
\bibitem{antal2} Antal T, Krapivsky PL, Redner S. Social balance on networks: The dynamics of friendship and enmity. Physica D: Nonlinear Phenomena. 2006; 224(1-2):130–136. \url{https://doi.org/10.1016/j.physd.2006.09.028}
\bibitem{Marvel1} Marvel SA, Kleinberg J, Kleinberg RD, Strogatz SH. Continuous-time model of structural balance. PNAS. 2011; 108(5), 1771–1776. \url{https://doi.org/10.1073/pnas.1013213108} PMID:\href{https://pubmed.ncbi.nlm.nih.gov/21199953/}{21199953}
\bibitem{Marvel2} Marvel SA, Strogatz SH, Kleinberg JM. Energy Landscape of Social Balance. Physical Review Letter. 2009;
103:198701.  \url{https://doi.org/10.1103/PhysRevLett.103.198701} PMID:  \href{https://pubmed.ncbi.nlm.nih.gov/20365960/}{20365960}
\bibitem{Abell} Abell P, Ludwig M. Structural Balance: A Dynamic Perspective. The Journal of Mathematical Sociology.
2009; 33(2):129–155. \url{https://doi.org/10.1080/00222500902718239}
\bibitem{Traag} Traag VA, Van Dooren P, De Leenheer P. Dynamical Models Explaining Social Balance and Evolution
of Cooperation. PLOS ONE. 2013; 8(4):1–7. \url{https://doi.org/10.1371/journal.pone.0060063} PMID:\href{https://pubmed.ncbi.nlm.nih.gov/23634204/}{PMC3636264}
\bibitem{Gawronski} Gawronski P, Gronek P, Kulakowski K. The Heider balance and social distance. arXiv:physics/0501160 [Preprint]. 2005 [cited 2005 Jan 31]. Available from: \url{https://arxiv.org/abs/physics/0501160}
\bibitem{Hedayatifar} Hedayatifar L, Hassanibesheli F, Shirazi A, Farahani SV, Jafari GR. Pseudo paths towards minimum energy states in network dynamics. Physica A. 2017; 483(7307), 109–116. \url{https://doi.org/10.1016/j.physa.2017.04.132}
\bibitem{Rabbani} Rabbani F, Shirazi AH, Jafari GR. Mean-field solution of structural balance dynamics in nonzero temperature. Physical Review E. 2019; 99, 062302. \url{https://doi.org/10.1103/PhysRevE.99.062302}
\bibitem{Castellano}  Castellano C, Fortunato S, Loreto V. Statistical physics of social dynamics. Reviews of Modern Physics. 2009; 81(2), 591. \url{https://doi.org/10.1103/RevModPhys.81.591}
\bibitem{borjiorno} Bongiorno C, Challet D. Nonparamrtric sign prediction of high-dimensional correlation matrix coefficients. 
EPL. 2021; 133(4), 48001. \url{https://doi.org/10.1209/0295-5075/133/48001}
\bibitem{Kulakowski} Kulakowski K, Gawronski P, Gronek P. The Heider Balance: A Continuous Approach. International Journal of Modern Physics C. 2005; 16(5)707-716. \url{https://doi.org/10.1142/S012918310500742X}
\bibitem{Krawczyk}  Krawczyk MJ, Wołoszyn M, Gronek P, Kułakowski K, Mucha J. The Heider balance and the looking-glass self: modelling dynamics of social relations. Scientific Reports. 2019; 9, 1-8. \url{https://doi.org/10.1038/s41598-019-47697-1} PMID:\href{https://pubmed.ncbi.nlm.nih.gov/31371775/}{PMC6671965}
\bibitem{harrary} Harary F, Kabell JA. A simple algorithm to detect balance in signed graphs. Mathematical Social Sciences. 1980; 1(1), 131–136. \url{https://doi.org/10.1016/0165-4896(80)90010-4}
\bibitem{Rijt} Rijt AVD. The micro-macro link for the theory of structural balance. Journal of Mathematical Sociology. 2011; 35(1-3), 94–113.\url{https://doi.org/10.1080/0022250X.2010.532262}
\bibitem{Kargaran} Kargaran A, Ebrahimi M, Riazi M, Hosseiny A,  Jafari GR. Quartic Balance Theory: Global Minimum With Imbalanced Triangles. Physical Review E. 2020; 102, 012310. \url{https://doi.org/10.1103/PhysRevE.102.012310}
\bibitem{Masoumi} Masoumi R, Oloomi F, Kargaran A, Hosseiny A, Jafari GR. Mean-field solution for critical behavior of signed networks in competitive balance theory. Physical Review E. 2021; 103, 052301 .\url{https://doi.org/10.1103/PhysRevE.103.052301}
\bibitem{Oloomi}  Oloomi F, Masoumi R, Karimipour K, Hosseiny A, Jafari GR. Competitive balance theory: Modeling conflict of interest in a heterogeneous network. Physical Review E. 2021; 103, 022307 .\url{https://doi.org/10.1103/PhysRevE.103.022307}
\bibitem{Hart}  Hart J. Symmetry and polarization in the European international system, 1870-1879: a methodological
study. Journal of Peace Research. 1974; 11(3):229–244. \url{https://doi.org/10.1177/002234337401100307}
\bibitem{Hummon}  Hummon NP, Doreian P. Some Dynamics of Social Balance Processes: Bringing Heider back into Balance Theory. Social Networks. 2003; 25(1):17–49. \url{https://doi.org/10.1016/S0378-8733(02)00019-9}
\bibitem{Doreian} Doreian P, Mrvar A. Structural Balance and signed international relations. Journal of Social Structure. 2015; 16(1).\url{https://doi.org/10.21307/joss-2019-012}
\bibitem{Lerner} Lerner J. Structural balance in signed networks: Separating the probability to interact from the tendency
to fight. Social Networks. 2016; 45:66–77. \url{https://doi.org/10.1016/j.socnet.2015.12.002}
\bibitem{Saiz}Saiz H, Gómez-Gardeñes J, Nuche P, Girón A, Pueyo Y, Alados CL. Evidence of structural balance in
spatial ecological networks. Ecography. 2017; 40(6):733–741. \url{https://doi.org/10.1111/ecog.02561}
\bibitem{Zheng} Zheng X, Zeng D, Wang FY. Social balance in signed networks. Information Systems Frontiers. 2015; 17,1077–1095. \url{https://doi.org/10.1007/s10796-014-9483-8}
\bibitem{Ilany} Ilany A, Barocas A, Koren L, Kam M, Geffen E. Structural balance in the social networks of a wild mammal. Animal Behaviour. 2013; 85(6), 1397–1405. \url{https://doi.org/10.1016/j.anbehav.2013.03.032}
\bibitem{Moradimanesh}  Moradimanesh Z, Khosrowabadi R, Eshaghi Gordji M, Jafari GR. Altered structural balance of resting-state networks in autism. Scientific Reports. 2021; 11:1966 \url{https://doi.org/10.1038/s41598-020-80330-0}
\bibitem{Davis} Davis JA. Clustering and Structural Balance in Graphs. Human Relations-SAGE. 1967; 20, 181. \url{https://doi.org/10.1177/001872676702000206}
\bibitem{Acharya} Acharya BD. Spectral criterion for cycle balance in networks. Journal of Graph Theory. 1980; 4(1), 1-11. \url{https://doi.org/10.1002/jgt.3190040102}
\bibitem{Srinivasan} Srinivasan A. Local balancing influences global structure in social networks. PNAS. 2011; 108(5), 1751–1752. \url{https://doi.org/10.1073/pnas.1018901108}
\bibitem{Estrada} Estrada E, Benzi M. Walk-based measure of balance in signed networks: Detecting lack of balance in social networks. Physical Review E. 2014; 90:042802. \url{https://doi.org/10.1103/PhysRevE.90.042802} PMCID:\href{https://pubmed.ncbi.nlm.nih.gov/25375544/}{25375544}
\bibitem{Costanzo} Costanzo M, et al. A global genetic interaction network maps a wiring diagram of cellular function. Science. 2016; 353, 6306. \url{https://doi.org/10.1126/science.aaf1420} PMCID: \href{https://pubmed.ncbi.nlm.nih.gov/27708008/}{PMC5661885}
\bibitem{Costanzo2} Costanzo M, et al. The Genetic Landscape of a Cell. Science. 2010; 327, 425-431. \url{https://doi.org/10.1126/science.1180823} PMCID:\href{https://pubmed.ncbi.nlm.nih.gov/20093466/}{PMC5600254}
\bibitem{Blomenand} Blomen VA, et al. Gene essentiality and synthetic lethality in haploid human cells. Science. 2015; 350, 1092-1096. \url{https://doi.org/10.1126/science.aac7557} PMID: \href{https://pubmed.ncbi.nlm.nih.gov/26472760/}{26472760}
\bibitem{Winzeler} Winzeler EA, et al. 	Functional characterization of the S. cerevisiae genome by gene deletion and parallel analysis. Science. 1999; 7, 901–906. \url{https://doi.org/10.1126/science.285.5429.901} PMID: \href{https://pubmed.ncbi.nlm.nih.gov/10436161/}{10436161}
\bibitem{Allahyari} Allahyari N, Hosseyni A, Abedpour N, Jafari GR. Analyzing the heterogeneous structure of the genes interaction network through the random matrix theory. arXiv:2107.13300 [Preprint]. 2021 [cited 2021 Jul 28]. Available from: \url{https://arxiv.org/abs/2107.13300}
\bibitem{Parapouli} Parapouli M, Vasileiadis A, Afendra AS, Hatziloukas E. Saccharomyces cerevisiae and its industrial applications. AIMS Microbiol. 2006; 6, 1,1-31. \url{https://doi.org/10.3934/microbiol.2020001} PMCID:\href{https://pubmed.ncbi.nlm.nih.gov/32226912/}{32226912}
\bibitem{Baryshnikova} Baryshnikova A, et al. Quantitative analysis of fitness and genetic interactions in yeast on a genome scale. Nature Methods. 2010; 7, 1017–1024. \url{https://doi.org/10.1038/nmeth.1534} PMCID:\href{https://pubmed.ncbi.nlm.nih.gov/21076421/}{PMC3117325}
\bibitem{Li} Li Z, et al. Systematic exploration of essential yeast gene function with temperature-sensitive mutants. Nature Biotechnology. 2011; 29, 361-367.  \url{https://doi.org/10.1038/nbt.1832} PMID: \href{https://pubmed.ncbi.nlm.nih.gov/21441928/}{21441928}
\bibitem{newman} Newman MEJ. The structure and function of complex networks. SIAM Review. 2003; 45, 167–256. \url{https://doi.org/10.1137/S003614450342480}
\bibitem{Albert} Albert R, Barabási AL. Statistical mechanics of complex networks. Reviews of Modern Physics. 2002; 74, 47.  \url{https://doi.org/10.1103/RevModPhys.74.47}
\bibitem{Watts} Watts DJ, Strogatz SH. Collective dynamics of ‘small-world’ networks. Nature. 1998; 393 (6684) 440–442. \url{https://doi.org/10.1038/30918} PMID: \href{https://pubmed.ncbi.nlm.nih.gov/9623998/}{9623998}
\bibitem{Jalan} Jalan S, Solymosi N, Vattay G, Li B. Random matrix analysis of localization properties of gene coexpression network. Physical Review E. 2010; 81, 046118. \url{https://doi.org/10.1103/PhysRevE.81.046118}
\bibitem{Pradhan} Pradhan P, Jalan S. From spectra to localized networks: A reverse engineering approach. IEEE. 2020; 7, 4,3008-3017. \url{https://doi.org/10.1109/TNSE.2020.3008999}
\bibitem{Namaki} Namaki A, Raei R, Jafari GR. Comparing Tehran stock exchange as an emerging market with a mature market by random matrix approach. World Scientific Publishing Company. 2011; 8, 371-383. \url{https://doi.org/10.1142/S0129183111016300}
\bibitem{Saeedian} Saeedian M, Jamal T, Kamali MZ, Bayani H, Yasseri T, Jafari GR. Emergence of world-stock-market network. Physica A. 2019; 526, 120792 \url{https://doi.org/10.1016/j.physa.2019.04.028}
\bibitem{Terzi} Terzi E, Winkler M. A Spectral Algorithm for Computing Social Balance. In: Frieze A., Horn P., Prałat P. (eds) Algorithms and Models for the Web Graph. WAW 2011. Lecture Notes in Computer Science, Springer, Berlin, Heidelberg. 2011;6732. \url{https://doi.org/10.1007/978-3-642-21286-4_1}
\bibitem{somaye} Sheykhali S, Darooneh AH, Jafari GR. Partial balance in social networks with stubborn links. Physica A. 2020; 548, 123882. \url{https://doi.org/10.1016/j.physa.2019.123882}


\end{thebibliography}
\end{document}